\documentclass[float, aps, prf, showpacs, superscriptaddress, groupedaddress, letter, twocolumn, longbibliography]{revtex4-1}  

\usepackage{graphicx}  
\DeclareGraphicsExtensions{.png,.pdf,.eps}

\usepackage{dcolumn}   \usepackage{bm}        \usepackage{amsmath,amsfonts,amssymb}

\usepackage{xcolor}

\newcommand{\inprod}[2]{\left<{#1},{#2}\right>}

\def\x{\mathbf{x}}

\def\a{\mathbf{a}}
\def\u{\mathbf{u}}
\def\v{\mathbf{v}}

\def\f{\mathbf{f}}
\def\g{\mathbf{g}}

\def\Z{\mathbb{Z}}
\def\N{\mathbb{N}}

\def\S{{\mathrm{S}}}
\def\U{{\mathrm{U}}}

\def\R{{\mathbf{H}^m_\omega}}
\def\res{\mathbf{\psi}}
\def\for{\mathbf{\phi}}

\def\be{\begin{equation}}
\def\ee{\end{equation}}
\def\bea{\begin{eqnarray}}
\def\eea{\end{eqnarray}}

\newtheorem{theorem}{Theorem}       

\newtheorem{example}[theorem]{Example}
\def\bex{\begin{example}}
\def\eex{\end{example}}

\graphicspath{{figures/}{figures/channel/}}

\begin{document}

\author{Ati S. Sharma}
\email{a.sharma@soton.ac.uk}
\affiliation{University of Southampton, UK}

\author{Igor Mezi{\'c}}
\email{mezic@ucsb.edu}
\affiliation{University of California, Santa Barbara, USA}

\author{Beverley J. McKeon}
\email{mckeon@caltech.edu}
\affiliation{California Institute of Technology, USA}

\date{\today}

\title{On the correspondence between Koopman mode decomposition, resolvent mode decomposition, and invariant solutions of the Navier-Stokes equations.}
\begin{abstract}
                    
    The relationship between Koopman mode decomposition, resolvent mode decomposition and exact invariant solutions of the Navier-Stokes equations is clarified.
    The correspondence rests upon the invariance of the system operators under symmetry operations such as spatial translation.
    The usual interpretation of the Koopman operator is generalised to permit combinations of such operations, in addition to translation in time.
    This invariance is related to the spectrum of a spatio-temporal Koopman operator, which has a travelling wave interpretation.
        The relationship leads to a generalisation of dynamic mode decomposition, in which symmetry operations are applied to restrict the dynamic modes to span a subspace subject to those symmetries.
        The resolvent is interpreted as the mapping between the Koopman modes of the Reynolds stress divergence and the velocity field. It is shown that the singular vectors of the resolvent (the resolvent modes) are the optimal basis in which to express the velocity field Koopman modes where the latter are not \emph{a priori} known.
\end{abstract}


\maketitle

This paper presents a unifying view of a range of methods used to characterise
nonlinear solutions of the Navier-Stokes equations. Ordered by roughly
decreasing complexity in their treatment of nonlinearity, these
are invariant solutions, a generalised form of Koopman mode decomposition,
dynamic mode decomposition and resolvent mode decomposition.  The instances in
which the four methods of analysis coincide are identified here. We hope to
provide a more rigorous basis on which to interpret these methods and to inform the
potential user of the appropriate tool for a particular problem.

Invariant solutions are exact (nonlinear) solutions of the Navier-Stokes
equations that remain the same after certain symmetry operations are applied
(such as reflection, rotation or shifts in space or time). They are sometimes
called `exact coherent structures' and are implicated in the formation
of apparently repeating highly ordered spatio-temporal patterns in turbulence.
These solutions are the subject of an active and long-standing area of
research; see for example \cite{Kawahara.Uhlmann.van-Veen:2012,
Eckhardt.Schneider.Hof.ea:2007, Kerswell:2005} for reviews. From the viewpoint
of dynamical systems, turbulence may be understood in terms of a state
trajectory visiting the neighbourhood of various invariant solutions.

Koopman modes are a general way of analysing the dynamics of a nonlinear system
\cite{Mezic:2013, Budisic.Mohr.Mezic:2012, Mezic:2005,
Rowley.Mezic.Bagheri.ea:2009}. The Koopman modes arise from the spectral
analysis of the Koopman operator, which is an infinite-dimensional operator
that evolves functions of the system's state.  Koopman mode decomposition is
closely related to dynamic mode decomposition (DMD) \cite{
    Rowley.Mezic.Bagheri.ea:2009, Schmid.Sesterhenn:2008, Schmid:2010,
Wynn.Pearson.Ganapathisubramani.ea:2013, Kutz.Brunton.Luchtenburg.ea:2014,
Jovanovic.Schmid.Nichols:2014}, which is a popular way to decompose data into
time-varying modes, and DMD modes approximate Koopman modes in certain
circumstances.

Independently, the resolvent mode decomposition \cite{McKeon.Sharma:2010} arose
as an analysis of wall turbulence that provides an ordered orthonormal basis to
express turbulent flow fields in an efficient and physically meaningful way.
Like invariant solutions, it has been used to explain the phenomenon of
coherent structure \cite{Sharma.McKeon:2013}, which raised the possibility that
the two apparently disjoint approaches are actually related.  Following that
work, it has been shown that the resolvent mode basis can be used to
efficiently capture many nonlinear invariant solutions of the Navier-Stokes
equations \cite{Sharma.Moarref.McKeon.ea:2016}.

In the following, we clarify the fundamental relationship between these various
different approaches.  The relationship rests on the idea that Koopman operator
theory can be generalised to consider group actions such as spatial
translations in addition to translation in time.  We will show that each
invariant solution satisfying a set of symmetries is associated with an
eigenvalue problem for a Koopman operator defined by the same symmetries. For
the case of spatial and temporal translation, the resulting Koopman modes have
the natural interpretation of travelling waves.  We present a similar
generalisation of DMD that provides a simple way of computing a basis for the
same subspace from experimental or simulation data. During the late stages of
preparation of this manuscript, a related idea of DMD in a moving frame of
reference was independently proposed by Sesterhenn and Sharipour
\cite{Sesterhenn.Shahirpour:2016}. Here we provide a theoretical basis for that
idea.  It has previously been shown that the spatial invariance of a linear system
is inherited by distributed linear optimal control of that system and that the
control is block-diagonalised by the Fourier transform
\cite{Bamieh.Paganini.Dahleh:2002}. The present result is comparable, but
applies in the fully nonlinear setting.

It is also shown that the resolvent operator arising from the Navier-Stokes
equations can be interpreted as a mapping between two observables (from
nonlinear forcing to velocity). As a consequence, the resolvent modes provide a
natural basis in which to expand the Koopman modes. Often, a projection of the
velocity field onto the leading resolvent modes will well approximate the
Koopman modes, in which case the dynamics are essentially low-dimensional. The
resolvent modes coincide with the spectral decomposition of the Koopman
operator in all invariant directions. Since the resolvent modes can be
calculated from only the Navier-Stokes equations and the time-space mean
velocity field, the resolvent modes offer a useful proxy for Koopman modes when
calculation of DMD modes is impractical or impossible.  In addition, because
travelling wave invariant solutions are also economically expressed as a
Koopman mode expansion, this also explains why projections of invariant
solutions onto their resolvent modes seem to approximate invariant solutions so
well.

This line of thought also suggests that it is reasonable to
approximate the dynamics of turbulent flow having a continuous
temporal spectrum with a projection onto a discrete set of
frequencies; this is entirely analogous to a periodic flow domain
being used to approximate an infinite one.

A brief introduction to Koopman modes will be given in the following.  For a
more comprehensive introduction to Koopman modes we refer the reader to other
sources \cite{Mezic:2013, Budisic.Mohr.Mezic:2012, Mezic:2005, Rowley.Mezic.Bagheri.ea:2009}.
Consider the case where dynamics of a time-varying vector-valued field on a one-dimensional spatial domain $\u(x,t)$ are governed by continuous dynamics,
\begin{equation}
    \partial_t \u(x,t) = \f\left(\u(x,t)\right)
    \label{dynamics}
\end{equation}
with observations given by $\g(\u(x,t))$.
All vectors are written in bold font (e.g.: ${\bf u}$).

For simplicity, the technical development is specialised to a system following a periodic orbit, resulting in a discrete spectrum of the Koopman operator, but the results are expected to generalise in a straightforward way to systems with continuous spectra, including turbulence.

Define a temporal shift operator,
\begin{equation}
    \S^{t'} \u(x,t) = \u(x,t+t').
\end{equation}
A family of Koopman operators
$\U^{t}$ \cite{Koopman:1931, Lasota.Mackey:1994} is defined for an arbitrary $\u(x,t)$ and for an arbitrary function $h$ via
\begin{equation}
    \U^{t'} h(\u(x,t)) = h(\S^{t'} \u(x,t)) = h(\u(x,t+t')).
\end{equation}
It is instructive to consider the spectral properties of the Koopman operator \cite{Mezic:2005},
\begin{equation}
    \U^{t'} \varphi_n(\u(x,t)) = \mu_n^{t'} \varphi_n(\u(x,t)).
\end{equation}
Consider the observable $\g$ defined earlier. Since we have assumed the case of a period orbit, we may expand the observable in terms of the eigenfunctions of $\U^{t}$,
\begin{equation}
    \g(\u(x,t)) = \sum_{n \in \Z} \varphi_n (\u(x,t)) \g_n(x).
\end{equation}
The $\g_n$ are the projections of the observable $\g$ onto the eigenfunctions of $\U^{t}$, and are called  Koopman modes.
For simplicity of presentation, we shall take $\g=I$ in what follows.
Then, expanding $\u(x,t)$ accordingly, we find that
\begin{equation}
    \u(x,t) = \sum_{n\in\Z}  \mu_n^{t} \varphi_n \u_n(x),
\end{equation}
where $\u_n(x)$ is the $n$-th Koopman mode. Given a sufficiently long time for the state to decay onto the attractor, and the absence of a continuous spectrum, all $\mu_n^t$ in the expansion will be the temporal Fourier modes.
Indeed, the eigenvalues of the operator are $\mu_n^{t'} = e^{in\omega t'}$ and $\U^t$ becomes unitary \cite{Mezic:2005}.
In general, eigenvalues of $\U^t$ can be off the unit circle, for example decaying solutions of fixed spatial shape are associated with a spectral radius of less than unity. It is straightforward to generalise the previous discussion to this case.
The limit cycle can therefore be decomposed in terms of these eigenfunctions and Koopman modes,
\begin{equation}
    \u(x,t) = \u^*(x) + \sum_{n\in\Z\setminus 0} \u_n(x) e^{in\omega t}.
    \label{temporal Koopman decomposition}
\end{equation}
The Koopman mode $\u_n(x)$ is the projection of the field $\u(x,t)$ on the subspace spanned by an eigenfunction of $\U^t$ \cite{Mezic:2013}.
Notice that the temporal average emerges as $\u^*(x)$. In the case of the continuous spectrum, rather than coefficients over discrete frequencies, we must use densities over a continuum.
The interpretation is that an expansion in neutral modes is most natural once the system has decayed onto the attractor.

Next we will show that travelling waves and periodic orbits arise from the eigenvalue problem for a family of joint spatio-temporal Koopman operators.
We earlier defined the Koopman operator with respect to a time-shift.
Here we introduce the spatial analogue to the Koopman operator, which we will call the spatial Koopman operator.
Similarly to the temporal case described earlier, we begin by defining a composition operator $\U^{x'}$ via
\begin{equation}
    \U^{x'} h(\u(x, t)) = h(\u(x+x', t)).
\end{equation}
Again, assuming periodicity in $x$, eigenvalues of $\U^{x'}$ associated with Koopman modes present in the flow lie on the unit circle. As such, $\U^{x'}$ will be unitary in this case. Since unitary operators are norm-preserving, again, the interpretation is that an expansion in neutral modes is most natural once the system has reached statistical homogeneity in $x$.
This is essentially the same situation as the temporal Koopman operator except with the shift being in $x$ rather than $t$.
In the more general case of a spatially developing flow, $\U^{x'}$ will have eigenvalues off the unit circle associated with the spatial growth or decay of the corresponding spatial Koopman modes.

Similarly to $\U^{t}$, the operator may have a continuous and point spectrum. Restricting our attention to the spatially periodic case (the spatial analogue to limit cycle behaviour), we may expand with just a point spectrum as
\begin{equation}
    \u(x,t) = \bar{\u}(t) + \sum_{l\in\Z\setminus 0} \u_l(t) e^{il\alpha x}
\end{equation}
with $\alpha$ the fundamental wavenumber.
The spatial Koopman modes are $\u_l(t)$ and this time, the spatial average appears as $\bar{\u}(t)$.

Having defined a spatial Koopman operator, it is natural to wonder what happens when the spatial and temporal Koopman operators are combined. To find out, we define a spatio-temporal Koopman operator,
\begin{align}
    \U^{x',t'} h(\u(x, t)) =& \U^{x'} \U^{t'} h(\u(x, t)) \nonumber \\
    =& h(\u(x + x', t + t')).
\end{align}

Consider the eigenvalues and eigenvectors of the compound or spatio-temporal Koopman operator,
\begin{equation}
    \U^{x'} \U^{t'} \varphi_j(\u(x,t)) = \mu_j^{x',t'} \varphi_j (\u(x,t)).
\end{equation}
These will be functions that are the same up to a scalar factor after translation by $t'$ then by $x'$ (later, but downstream).
This arrangement is equivalent to defining a single Koopman operator for the combined transformation.

The $\mu_j^{x,t}$ that satisfy this are $e^{i(n\omega t + l\alpha x)}$ giving the expansion in spatio-temporal Koopman modes,
\begin{equation}
    \u(x,t) = \sum_{n,l\in\Z} \u_{n,l} e^{i (n \omega t + l \alpha x)},
    \label{tw-expansion}
\end{equation}
with $\alpha$ the fundamental wavenumber and $\omega$ the fundamental frequency.
The obvious interpretation is an expansion in travelling waves with downstream phase speed $c=n\omega / l \alpha$.
This goes some way to explaining how a spatial periodicity induces a sparsity in temporal frequency in turbulent simulations \cite{Gomez.Blackburn.Rudman.ea:2014, Bourguignon.Tropp.Sharma.ea:2014}, by restricting the $\omega$ available to satisfy a range of wavespeeds.
Again, this is easily generalisable to systems that develop spatially, in which case the eigenvalues of $\U^x$ will be off the unit circle.

Notice that the spatio-temporal mean occurs naturally, $\u_{0,0} = \bar{\u}^*$.
Clearly, $\U^{x,t} = \U^{t,x}$.
In this sense, the expansion in travelling waves is natural and arises directly from the spatio-temporal symmetries of the equations determining the dynamics.
This idea may be generalised further to include other operations, such as reflection (decomposing into odd and even functions), or indeed any other symmetries induced by group action.

Comoving frames of reference are well studied in the context of invariant solutions (\cite{Kreilos.Zammert.Eckhardt:2014, Willis.Cvitanovic.Avila:2013} for example) and so unsurprisingly this interpretation is reminiscent of the formulation used to specify exact travelling wave solutions, which are fixed points in a comoving frame of reference. Let $\v(x,t)$ be such a solution to \eqref{dynamics} that satisfies
\begin{equation}
    \v(x,t) = \v(x + \delta x, t + \delta t),
\end{equation}
with $c=\delta x / \delta t$ being its phase speed.
Then, the decomposition induced by the spatio-temporal Koopman operator, according to \eqref{tw-expansion}, is
\begin{equation}
    \v(x,t) = \sum_{n,l\in\Z} \v_{n,l} e^{i (n \omega t + l \alpha x)}.
\end{equation}
Since the travelling wave has a single wave speed $c$, we conclude that all $\v_{n,l}$ are zero except those where $n\omega / l \alpha = \delta x / \delta t = c$.
From this we can see that the spatio-temporal Koopman modes are a natural and parsimonious way to express travelling wave solutions.

This analysis suggests a new generalisation of DMD, which is described as follows.
Usually, the DMD of a dataset proceeds as follows. First, a series of velocity field snapshots are taken and formed into a matrix,
\begin{equation}
    Z_t = \left[ \u_1 \  \u_2 \  \ldots \  \u_{N-1} \right].
\end{equation}
A second snapshot matrix is formed from the same data,
\begin{equation}
    Z_{t+\Delta t} = \left[ \u_2 \  \u_3 \  \ldots \  \u_{N} \right],
\end{equation}
where a snapshot $\u_{i+1}$ is $\Delta t$ later than the snapshot at $\u_i$. While $\u_{i+1}$ is generated by the nonlinear dynamics from $\u_i$, DMD approximates this behaviour by a linear operator, and the following problem is formed,
\begin{equation}
    Z_{t+\Delta t} = M Z_{t}.
\end{equation}
The DMD modes are the eigenvectors of $M$, which may be found by various methods
\cite{Rowley.Mezic.Bagheri.ea:2009, Schmid:2010, Wynn.Pearson.Ganapathisubramani.ea:2013, Kutz.Brunton.Luchtenburg.ea:2014, Jovanovic.Schmid.Nichols:2014}.

Another way to look at this problem is to introduce a time-translation operator acting on the matrix of snapshots,
\begin{equation}
    \tau_{\Delta t} Z_{t} \equiv Z_{t+\Delta t}= M Z_{t}.
\end{equation}
It is clear from the earlier discussion that the DMD problem more generally applies to any combination of operators, such as reflection in $z$ ($\sigma_z$), translation in the streamwise direction ($\tau_{\Delta x}$), or time. The original paper introducing DMD \cite{Schmid:2010} did present a generalisation to spatially developing flows, but did not discuss the combination with time-shift or other transformations such as reflection.

Using the interpretation of DMD modes as providing a basis with which to approximate the mapping between two observables, we may do the same with time- and space-shifted snapshot matrices.
More generally, by applying a transformation (in this case, translation in space and time), the DMD modes become the Koopman modes in the associated Koopman mode expansion (e.g.~\eqref{tw-expansion}).

For example, the DMD problem
\begin{equation}
    \tau_{\Delta x} \tau_{\Delta t} \sigma_z Z_{t,x} \equiv Z_{t+\Delta t, x+\Delta x, \sigma_z}= M Z_{t,x}
\end{equation}
defines the linear operator $M$ that approximates the nonlinear operation that relates snapshots reflected in $z$, moved downstream by $\Delta x$ and later by $\Delta t$ to the current snapshot. As such the eigenvectors of $M$ (the DMD modes) will all be travelling waves, symmetric or anti-symmetric in $z$ and convecting with speed according to \eqref{tw-expansion}.

The eigenvalues are interpreted in the usual way, but in the moving frame of reference.
We may solve for eigenvectors of $M$ with the usual DMD methods.
Since this subspace is smaller than the space of regular DMD modes, we might expect such symmetrised DMD modes to converge faster. In exploratory computations, we have observed this to be the case.
For data periodic in $x$, the eigenvalue problem for $\U^x$ or $M$ will reduce to a discrete Fourier transform in $x$, in much the same way as happens with $\U^t$.

The decomposition into travelling waves in spatially invariant directions leaves the question of how to decompose non-invariant directions. In the following, we specialise on the case of incompressible fluid flow.
The dynamics of incompressible fluid flow is most commonly studied using the Eulerian framework, via the
 non-dimensional version of Navier-Stokes equations
\begin{equation}
    \label{NS}
    \partial_t \u + {\bf u} \cdot \nabla {\bf u} = - \nabla p + \frac{1}{Re} \nabla^2 {\bf u}; \quad \nabla \cdot {\bf u} = 0,
\end{equation}
where $Re = \rho U L/ \mu$ is the Reynolds number,  $U$ a characteristic velocity, $L$ a length scale, $T$ a time scale and $\nabla$ is the spatial gradient operator. The density of the fluid is $\rho$ and its viscosity $\mu$. Velocity and pressure are non-dimensional, where velocity is scaled by $U$ and pressure is scaled by $\rho U^2$.

Assume a flow with a limit cycle of frequency $\omega$. We first consider the temporal Koopman mode decomposition for $\u$, as in \eqref{temporal Koopman decomposition},
\begin{equation}
    \u(\x,t)=\u^*(\x)+\sum_{n\neq 0}e^{in\omega t}\u_n(\x),
\end{equation}
where $\u^*(\x)$ is the time-averaged velocity. Denote $\tilde\u(\x,t)=\u(\x,t)-\u^*(\x)$.
Now, expand the pressure in its own Koopman modes:
\[
    p(\x,t)=p^*(\x)+\sum_{n\in\Z}e^{in\omega t}p_n(\x).
\]
In addition, expand $\tilde\u(\x,t)\cdot \nabla \tilde\u(\x,t)$ in its Koopman modes
\begin{equation}
    \tilde\u(\x,t)\cdot \nabla \tilde\u(\x,t)=\sum_{n\in \Z}e^{in\omega t}\f_n(\x),
    \label{nonlin}
\end{equation}
where the summation is over $n\in \Z$ since the time-average of $\tilde \u\cdot \nabla \tilde\u$ does not to have to be zero,
even though the time average of $\tilde \u$ is.
Essentially, we are treating the nonlinear term as a separate observable.

We now rewrite the Navier Stokes equation as
\begin{align}
    \label{eq:KoopNS}
    \sum_{n\neq 0} & in\omega e^{in\omega t}\u_n +  {\bf u^*} \cdot \nabla { \u^*}   + {\bf u^*} \cdot \nabla {\tilde \u}+ {\tilde \u} \cdot \nabla {\u^*} \nonumber \\
    =& - \nabla\left(p^*+\sum_{n\in\Z}e^{in\omega t}p_n \right) \nonumber \\
    +& \frac{1}{Re} \nabla^2\left( \u^*+\sum_{n\neq 0}e^{in\omega t}\u_n\right) - {\tilde \u} \cdot \nabla {\tilde \u},
     \nonumber \\
\end{align}
Assuming we know $\u^*$, and
integrating (\ref{eq:KoopNS}) against $e^{-i\omega mt}$, we get
\begin{align}
    im\omega \u_m + {\bf u^*} \cdot \nabla { \u_m}+ { \u_m} \cdot \nabla {\u^*} \\
    = - \nabla p_m + \frac{1}{Re} \nabla^2 \u_m- \f_m. \nonumber
\end{align}
The $n=0$ equation clearly reads
\begin{equation}
    {\bf u^*} \cdot \nabla { \u^*}=- \nabla p^* + \frac{1}{Re} \nabla^2\u^* - \f_0.
\end{equation}
Now we discuss the pressure terms. Because of incompressibility, we have
\[
    -\nabla^2 p=\nabla\cdot (\u\cdot\nabla\u).
\]
Thus, clearly,
\[
    -\nabla^2 p^*=\nabla(\u^*\cdot\nabla\u^*) + \nabla\cdot \f_0
\]
For higher order Koopman modes,
\begin{equation}
    \nabla^2 p_m=-\nabla\cdot (\u^*\cdot\nabla\u_m)-\nabla\cdot (\u_m\cdot\nabla\u^*)-\nabla \cdot \f_m.
\end{equation}
Thus, $p_m$ is related to $\u_m$ and $\f_m$ via a linear operator. Let us denote
\[
    L_1(\u_m)=(\nabla^2)^{-1}\left(-\nabla\cdot (\u^*\cdot\nabla)\u_m- \nabla\cdot (\u_m\cdot\nabla\u^*)\right)
\]
(taking care with boundary conditions when inverting the Laplacian)
and
\[
    L_2(\f_m)=-(\nabla^2)^{-1}\nabla \cdot \f_m.
\]
We get
\[
    p_m = L_1(\u_m) + L_2(\f_m).
\]
Note that if we define $L(\u^*)=\u^*\cdot\nabla+(\nabla \u^*)^\top$ and noticing $\Pi = (I - \nabla(\nabla^2)^{-1}\nabla\cdot)$ is the Leray projection, we get
\begin{equation}
    \left(im\omega I + \Pi L(\u^*) - \frac{1}{Re} \nabla^2\right)\u_m
    = -\Pi \f_m.
\end{equation}
Naming
\begin{equation}
    \R=-\left(im\omega I + \Pi L(\u^*) - \frac{1}{Re} \nabla^2\right)^{-1}\Pi
\end{equation}
the resolvent operator, we have obtained a relationship between the Koopman modes of $\tilde\u \cdot \nabla \tilde\u$
and Koopman modes of $\u$ via $\R$ as
\begin{equation}
    \u_m = \R\f_m.
    \label{resolvent}
\end{equation}
Notice that $\R$ depends on the time-average, which appears entirely naturally and with a consistent interpretation. The importance of the spectrum of the linear operator about $\u^*$ thus has a clear interpretation even in a fully nonlinear flow.

We would like to find a sensible basis in which to expand $\u_m$.
Ideally, the functions should be orthonormal, and chosen and
ordered in such a way that a truncation of the expansion should
still approximate the true $\u_m$ in a quantifiable way.
In the case that the dynamics are translation-invariant, we have already seen that the spatial Koopman modes are the correct choice. If this is not the case, the choice is more delicate.

To proceed, notice that $\R$ is a linear mapping between two observable fields. So it is reasonable to use two different bases for $\u_m$ and $\f_m$.
One sensible choice is the Schmidt decomposition of $\R$, because the truncation in that basis is optimal in the Frobenius norm.
The Schmidt decomposition is the infinite-dimensional equivalent to the singular value decomposition \cite{Young:1988}.
Then,
\begin{align}
    \R \f_m(\x) = \sum_{j\in \N} \sigma_j^m \inprod{\f_m(\x)}{\for_j^{m}(\x)} \res_j^m(\x) ,\\
    \inprod{\res_j^m(\x)}{\res_{j'}^m(\x)}=\delta_{j, j'},\nonumber \\
    \inprod{\for_j^m(\x)}{\for_{j'}^m(\x)}=\delta_{j, j'}\nonumber,\\
    \sigma_j^m \geq \sigma_{j+1}^m.
    \label{svd}
\end{align}
The pairs $\phi_j^m$ and $\psi_j^m$ are the Schmidt pairs (singular vectors) in the decomposition. The sets of $\phi_j^m$ and $\psi_j^m$ each form an orthonormal basis, with basis functions ordered by the singular values $\sigma_j^m$. This ordering provides a criterion for truncation.
In fact, $\R$ often has very large separation between the leading (one or two) singular values and the next. The physical basis for this is well documented in previous work \cite{McKeon.Sharma:2010}. It also turns out that in any shear flow, $\R$ is non-normal, so in general $\res_j^m(\x) \neq \for_j^m(\x)$.
This decomposition is like proper orthogonal decomposition \cite{Holmes.Lumley.Berkooz:1996}, but on a dynamical flow operator instead of a dataset.

This decomposition leads naturally to ordered expansions for both $\u_m$ and $\f_m$,
\begin{align}
    \u_m(\x) = \sum_{j\in \N} \chi_j^m \sigma_j^m \res_j^m (\x), \\
    \f_m(\x) = \sum_{j\in \N} \chi_j^m \for_j^m (\x).
\end{align}
In previous work we have called the set of $\res_j^m$ the response modes and the set of $\for_j^m$ the forcing modes.

If it so happens that $\sigma_1\gg \sigma_2$, we may reasonably approximate the Koopman mode $\u_m$ (up to a complex coefficient) by $\u_m(\x) \cong \res_1^m(\x)$, regardless of our knowledge of $\f_m$ (the argument does not hold the other way; to approximate $\f_m$ it would require $(\R)^{-1}$ to be approximately rank-1, which it is not).
The phrase `up to a complex coefficient' means that, while the functional form of each response mode is known, the phase and magnitude (making up the complex coefficient of the response mode) is not determined by the decomposition.
Similarly, Koopman modes are always defined up to a complex coefficients since they are projections onto eigenfunctions, which themselves are always defined up to a complex coefficient. In travelling wave terms, this means that the wave is defined up to an arbitrary phase, which is fixed in the coefficient.

However, because the \emph{relative} phase between forcing and response mode pairs is fixed by the decomposition, the phase between different response modes may be fixed either via calculation of the nonlinear forcing, by a projection onto Koopman modes determined by DMD, by fitting to a limited set of measurements \cite{Gomez.Blackburn.Rudman.ea:2016}, or by other methods \cite{Moarref.Jovanovic.Tropp.ea:2014, Moarref.Sharma.Tropp.ea:2013}.

In actuality, $\chi^m_j$ for leading few $j$ may be sufficiently small relative to the following coefficients that this effect may outweigh the effect of any separation of the leading singular values.
In such cases, it is reasonable to approximate $\R$ by a projection induced by \eqref{svd}, with the rank determined by the desired level of accuracy, as in \cite{Moarref.Jovanovic.Tropp.ea:2014}.
The extent to which either scenario applies will depend on the particularities of the system under study.

\begin{figure*}
    \centering
    \includegraphics[width=0.8\textwidth]{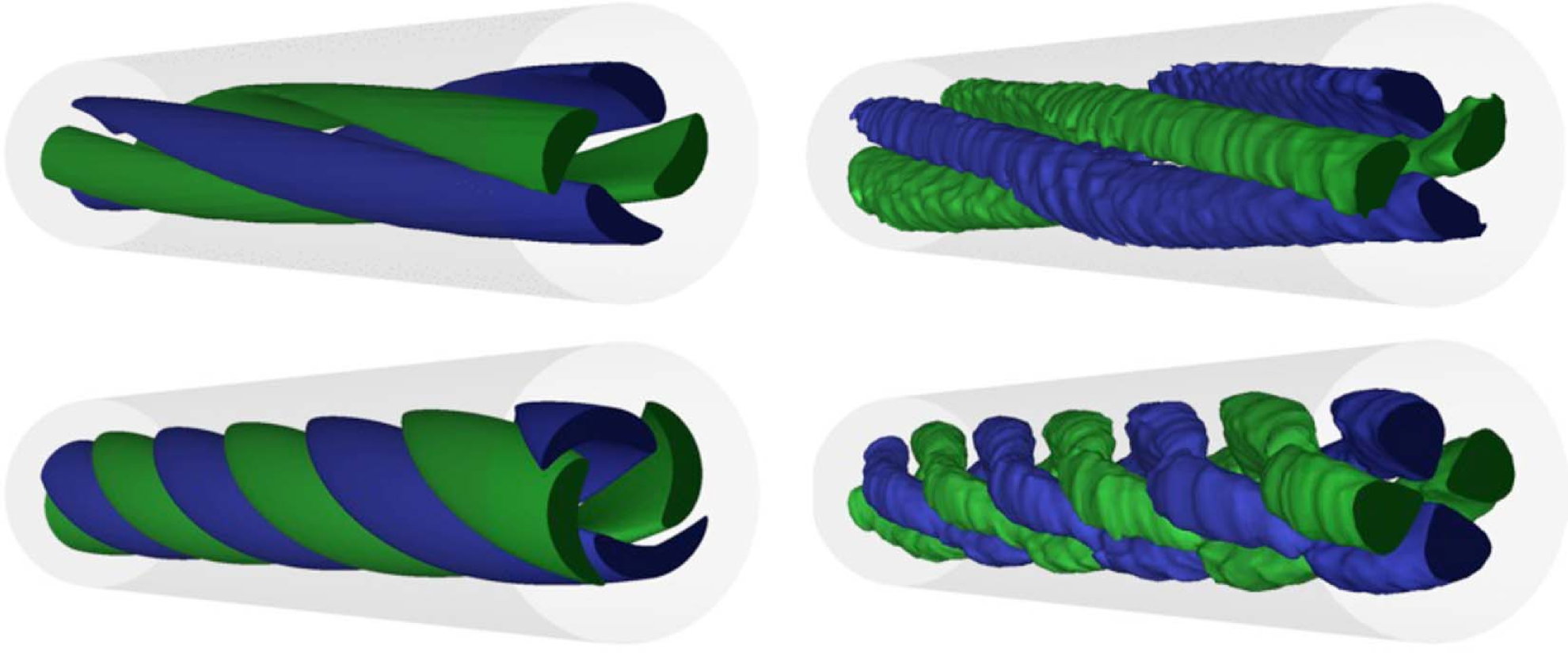}
    \caption{Modes in a turbulent pipe flow, length $2\pi$ diameters, $R^+=314$ based on friction velocity. Comparison between response modes (left) and DMD modes (right) at the same frequency ($\omega/2\pi = 0.1826$ (top) $\omega/2\pi = 0.5479$ (bottom)) with azimuthal wavenumber $2$. Coloured isosurfaces indicate $\pm 1/3$ of the maximum streamwise velocity fluctuation. Reproduced from G\'omez et al.~\cite{Gomez.Blackburn.Rudman.ea:2014}, with the permission of AIP Publishing.}
    \label{pipe-dmd}
\end{figure*}

We show a comparison of a response mode to a temporal DMD mode for turbulent flow in a pipe in Figure \ref{pipe-dmd}. Notice that in this case, as in all cases with spatial periodicity, spatio-temporal modes must also be modes of the temporal Koopman operator but with a spatial periodicity. In other words, although Figure \ref{pipe-dmd} was generated using a DMD of two-dimensional snapshots (at a single azimuthal Fourier wavenumber), it is clear from the previous analysis that the DMD/Koopman modes could equally have been found by using the Fourier transform in the streamwise direction, then by solving the much smaller resulting one-dimensional DMD problem.

In many canonical flows, such as flow through a pipe or channel, the flow equations are invariant under a spatial translation.
We will see next that the response modes coincide with the spectral decomposition of the Koopman operator in spatially invariant directions.
To avoid the inconvenience of treating a continuous spectrum, assume periodicity in $x$ and take $\x=(x,y)$.
Suppose we expand a (temporal) Koopman mode in terms of its response modes,
\begin{equation}
    \u_m(x,y) = \sum_{j\in \N} c_j^m \res_j^m(x,y).
\end{equation}
Applying the spatial Koopman operator gives (using its linearity)
\begin{equation}
    \U^{x} \u_m(x,y) = \sum_{j\in \N} c_j^m \U^{x} \res_j^m(x,y).
\end{equation}
The spatial Koopman mode expansion of $\u_m(x,y)$ is,
\begin{equation}
    \u_m(x,y) = \sum_{l \in \Z} \a_l(y) e^{il\alpha x}.
\end{equation}
Since $\res_j^m(x,y)$ form a complete orthonormal basis, satisfying the same symmetry property as the Koopman mode expansion, we have the result that
\begin{equation}
    c_j^m \res_j^m(x,y)=\a_l(y) e^{il\alpha x}
\end{equation}
where each $j$ is associated with a particular $l$.

This shows that decomposing into (i) the spatio-temporal Koopman modes then the response modes and (ii) temporal Koopman modes and then the response modes, are equivalent up to the ordering of the response modes.
That is, the spatial dependence on $x$ of $\res_j^m(x,y)$ will be the same as that of $\for_j^m(x,y)$ and the Koopman operator eigenvector of the same wavenumber where the resolvent is invariant under translation in $x$.

Such symmetry properties are intimately tied up with the normality of the resolvent and the availability of energy to flow perturbations. It is straightforward to show that the resolvent operator is normal with respect to the inner product over any invariant spatial direction. Since the mean profile enters the fluctuation equations and therefore the resolvent, the presence of mean shear in a direction breaks that invariance. This is manifested as the departure of the response modes from the Koopman modes in that spatial dependence.

The resolvent analysis has also been generalised to spatially developing flows by considering complex Fourier wavenumber \cite{McKeon.Sharma.Jacobi:2013}, in a way directly comparable to the Koopman modes.

We have shown that resolvent modes and exact solutions of the Navier-Stokes
equations share a common fundamental interpretation best expressed in terms of
the spectral properties of a class of Koopman operators. These Koopman
operators arise from symmetries of the governing equations (translation,
reflection and so on; for a complete discussion of the symmetries available in
Couette flow, for example, see \cite{Gibson.Halcrow.Cvitanovic:2009}). This
association has a natural physical interpretation as a decomposition into
travelling waves, and leads to dynamic mode decomposition for travelling waves.
It was shown that the application of these symmetries restricts the subspace
that the DMD modes lie in resulting in modes that obey these symmetries. We
expect this generalisation to prove useful for flows that develop
spatio-temporally, or where improved convergence is desired.

From this interpretation it becomes clear that, in principle, Koopman modes are
the `true' characterisation which DMD and resolvent modes may approximate.
However, DMD modes are empirically determined and the resolvent decomposition
is (almost entirely) analytical. Where the system is low-dimensional, it
follows that the spatial dependence of the Koopman modes may often be well
approximated by a small number of resolvent modes. This has the advantage that
parametric dependence may be cheaply explored (for determining the effects of
control, or for calculating Reynolds number scaling, for example).  Without
appeal to a nonlinear calculation or actual data, the resolvent mode
coefficients are not known. Therefore, where data is easily available, or where the
governing equations are not known, DMD modes may be the better choice, or some
combination of methods.  For example, in control problems, resolvent mode
coefficients can be fixed and the frequencies selected by DMD or another method
\cite{Gomez.Blackburn.Rudman.ea:2016}, while an expansion in resolvent modes can
inform how the true Koopman modes change as control is applied or boundary
conditions are changed \cite{Luhar.Sharma.McKeon:2014,
Luhar.Sharma.McKeon:2015}.

It has also been shown that the exact travelling wave solutions of the
Navier-Stokes equations have economical expansions using generalised Koopman
modes. This goes some way to explaining the (otherwise surprising) efficiency
of the resolvent modes in capturing fully nonlinear invariant solutions.

Although the presentation was specialised to systems with a discrete temporal
spectrum, the arguments should generalise to those with a continuous spectrum,
such as turbulence. In such systems, it seems reasonable to perform a
projection onto a discrete set of frequencies to approximate the original
system, by analogy with a periodic spatial flow domain being used to
approximate an infinite one.

The theory presented here has extensions for cases where other group actions and other spatial domains are considered (for instance, flows on a sphere).

\begin{acknowledgements}
    No significant data were generated for the purposes of EPSRC's data policy.
    IM was partially supported by the ARO contract W911NF-14-C-0102 and AFOSR FA9550-12-1-0230.
    BJM gratefully acknowledges the support of the ONR under grant N000141310739.
\end{acknowledgements}

\bibliography{refs}

\end{document}